*Article*

# Unlocking the Future of X-Ray Polarimetry with IXPE: Lessons Learned and Next Steps

Paolo Soffitta [1,*] , Enrico Costa [1] , Ettore Del Monte [1] , Alessandro Di Marco [1] , Sergio Fabiani [1] , Riccardo Ferrazzoli [1] , Fabio La Monaca [1] , Fabio Muleri [1] , Alda Rubini [1] and Alessio Trois [2]

1 INAF-IAPS, Via Fosso del Cavaliere 100, 00133 Rome, Italy; enrico.costa@inaf.it (E.C.); ettore.delmonte@inaf.it (E.D.M.); alessandro.dimarco@inaf.it (A.D.M.); sergio.fabiani@inaf.it (S.F.); riccardo.ferrazzoli@inaf.it (R.F.); fabio.lamonaca@inaf.it (F.L.M.); fabio.muleri@inaf.it (F.M.); alda.rubini@inaf.it (A.R.)
2 INFN-OAC, Via della Scienza 5, 09047 Selargius, Italy; alessio.trois@inaf.it
* Correspondence: paolo.soffitta@inaf.it

## Abstract

This paper discusses issues encountered during the early development of the instrument on the Imaging X-ray Polarimetry Explorer (IXPE), a NASA–ASI Small Explorer mission launched on 9 December 2021. IXPE has observed about 100 sources, yielding meaningful polarimetry for most of them. An on-board calibration system mitigated most non-ideal detector behaviors during operations. Data from the on-board polarized and unpolarized X-ray sources are routinely ingested by the flight pipeline to correct the instrument response in a manner transparent to users. Based on its scientific return and payload health, the IXPE mission has been extended through 2028. The lessons learned are informing the design of next-generation X-ray polarimetry missions, as discussed elsewhere in these conferences.

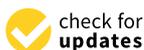

## 1. Introduction: The Imaging X-Ray Polarimetry Explorer in a Nutshell

The Imaging X-ray Polarimetry Explorer (IXPE) [1,2], a NASA–ASI SMEX mission currently operational, represents the culmination of decades of effort by a large community of scientists to follow-up the single positive OSO-8 measurement [3,4] with a more sensitive technique, thereby establishing this as observable as a routine tool in astronomy. In this paper, we emphasize the challenges surmounted during development, integration, and on-orbit operations, thanks to the dedication of an integrated Italy–US team. The Italian contribution includes ASI's programmatic and scientific management; the Space Science Data Center's provision of instrument flight pipeline modules; the Malindi Ground Station; and the leading roles of INAF and INFN in instrument developing, integration, testing, and calibration. On the industrial side, OHB-Italy served as a crucial contractor for the detector service unit and the design and development of the filter and calibration wheel. The U.S. team at NASA's Marshall Space Flight Center, the PI institution and provider of the mirrors and the Science Operations Center, worked in close partnership with the spacecraft developer (BAE, British Aerospace) and with the Laboratory for Atmospheric and Space Physics (LASP). This integrated team successfully addressed the challenges discussed in this paper, which we believe merit dissemination as lessons learned, four years after launch. In short, IXPE consists of three mirror modules [5], in which the instrument [2] hosts three detectors that each incorporate a Gas Pixel Detector (GPD) [6], which allows us to study

X-ray polarization from a celestial source by imaging the photoelectron tracks [7]. Scientists were aware of the importance of polarimetry for its ability to reveal emission processes and geometries in the vicinity of compact objects and for answering questions of fundamental physics in a completely novel way. This was made possible by the development of a dedicated ASIC [8]. IXPE was launched on 9 December 2021 from Cape Canaveral from the hystorical Pad 39A, and it detected polarization, observing about 100 different sources so far; see Table 1. filter and calibration wheels [9], one for each detector, are routinely used during Earth occultations to equalize the gain, verify the polarization sensitivity, and check for spurious polarization. The event data (photon-by-photon) and the ratemeter data are packetized by the Digital Service Units (DSUs) and stored in on-board memory for transmission to the ASI Ground Station at Malindi (Kenya). The Science Operations Center (SOC) at NASA's Marshall Space Flight Center (MSFC) in Huntsville organizes the long-term observing plan, manages Target-of-Opportunity (ToO) requests, and operates and maintains the flight pipeline, with the support of the ASI Space Science Data Center (SSDC). Data are received from the Mission Operations Center (MOC) at the Laboratory for Atmospheric and Space Physics (LASP), University of Colorado Boulder. HEASARC is responsible for the dissemination of data, making them publicly available, and issuing and managing the General Observer Program (GOP), which, for IXPE, is now at GOP number 3̃.

**Table 1.** The list of different sources observed by IXPE through the end of June 2025. Some of these sources were re-observed in order to improve the significance or to catch possible changes in the polarization properties in different states. PWNe = pulsar wind nebulae, SNRs = supernova remnants, BH = black holes, NSs = neutron stars, WDs = white dwarfs, AGN = active galactic nuclei.

|  | **Number of Objects** |
| --- | --- |
| 11 PWNe and isolated pulsars | Crab PWN, Vela PWN, MSH 15-52, PSR B0540-69, G21.5, 3C 58, PSR B1259-63, KES-75, PSR J1723-283.7, Lighthouse N, PSR J1023+0038 |
| 6 SNRs (7 pointings) | Cas A, Tycho's, NE SN 1006, RCW 86, RX J1713.7-3946, Vela Jr, SN1006SW |
| 17 Accreting stellar-BH | Cyg X-1, 4U 1630-472, Cyg X-3, LMC X-1, SS433, 4U 1957-115, SS 433 Lobe East, LMC X-3, SWIFT J1727.8-1613, 4U 1957+115, Swift J0243.6+6124, Swift J1727.8-1613, GX 339-4, SWIFT J151857.0-572, MAXI J1744-294, SS 433 Lobe West, GRS 1915+105 |
| 38 Accreting NS and WD | Cen X-3, Her X-1, GS1826-67, Vela X-1, Cyg X-2, GX 301-2, Xpersei, GX 9-9, 4U 1820, GRO J1008-57, XTE 1701-46, EXO 2030+375, LS V+44 17, GX 5-1, 4U 1624-49, Sco X-1, Cir X1, GX13+1, SMC X-1, SRGA J144459.2-604207, 4U 1538-52, V395 CAR, PSR J1023+00, GX 340+0, GX 3+1, 4U 1728-34, PSR J1723-2837, 4U 1735-44, GX 9+1, GX 349+2, 4U 1538-52, GX 17+2, 4U 1907+09, EX HYDRAE, 4U 1700-377, IGR J17091-3624, H 1417-624, UW CRB |
| 6 Magnetars | 4U 0142+61, 1RXS J170849, SGR 1806-20, 1E 2259+586, 1E 1841-045, 1E 1547.0-5408 |
| 10 Radio-quiet AGN and 1 Sgr A* molecular cloud | MCG 5-23-16, Circinus Galaxy, NGC 4151, IC 4329 A, Sgr A* Complex, NGC 1068, NGC 4945, NGC 2119, NGC 3227, 1ES 1927+654 |
| 21 Blazars and radio galaxies | Cen A, S5-0716-714, 1ES 1959-650, Mrk 421, BL Lac, 3C 454, 3C 273, 3C 279, Mrk 501,1ES 1959-650, BL-Lac, 1ES 0229-200, PG 1553 -113, S4 0954+65, 1E 2259+586, RGB J0710+591, H 1426+428, 1ES 1101-232, PICTOR A WEST, 1ES 1927+654,1ES 1927+654 |

## 2. The Gas Pixel Detector

The photoelectric effect in gas was proposed long ago as a method to measure the polarization of an X-ray photon beam either by searching for coincidences between the wires of a proportional counter [10] or by using the rise time in a single-wire proportional counter [11], and then with multi-wire proportional counters [12]. The direction of photoelectron emission has also been inferred by optically imaging the track [13] and, more recently, in a series of works [14–17]. Several authors attempted to exploit coincidences in the contiguous pixels of CCD [18–20] and CMOS sensors [21]. Track imaging with sufficient sensitivity became feasible thanks to advances in microelectronics in the late 1990s, enabling a new generation of gas detectors, either one-dimensional devices adapted from experiments at CERN [22,23] and later with instruments expressly developed to image the photoelectron tracks [6–8,24], for detecting polarization from celestial sources. These devices, known as Gas Pixel Detectors (GPDs) were developed by INFN and INAF-IAPS in Italy. They consist of a gas cell with a 1-cm thick drift gap bounded by a 50 μm aluminized beryllium X-ray window and the top metallic layer (copper) of a gas electron multiplier (GEM) [25,26]. The gas mixture is dimethyl ether (DME), filled to an initial pressure of 800 mbar (see Section 6), at Oxford X-ray Technologies OY, Espoo, Finland. The GEMs were produced by SciEnergy SciEnergy Co., Ltd in Japan (Japan). IXPE GEMs have 30 μm diameter holes, 50 μm pitch, and 50 μm thickness of the dielectric substrate (Liquid Crystal Polymer). An INFN-designed ASIC records the analog image of the photoelectron track with 105,600 hexagonally arranged pixels at a pitch of 50 μm. The ASIC implements self-triggering when a sufficient charge is collected in any pixel in a group of four, defining a region of interest (ROI) that includes a fiducial area of 20 pixels, and it also provides an external trigger output. With this information available, essentially all observables encoded in the X-ray beam—impact point, energy, arrival time, and linear polarization (but not circular polarization)—can be reconstructed by the analysis of the data provided by GPD.

## 3. An Extensible Boom Featuring a Tip–Tilt Rotation Mechanism But Without On-board Metrology System

IXPE was conceived and designed to be compatible with the Pegasus XL launcher in response to the call and Phase A requirements. Pegasus XL imposes stringent constraints on power, volume, and mass. Therefore, the only viable way to achieve the required focal length of 4 m was to adopt a deployable boom (mast). A similar approach had been used successfully before IXPE on the Nuclear Spectroscopic Telescope Array (NuSTAR), a NASA SMEX mission that remains operational.

Early IXPE designs included a NuSTAR-like metrology system, but this was removed during spacecraft integration for programmatic reasons. However, the optical system incorporated tip–tilt–rotation mechanisms to enable global post-deployment alignment of the optics; this mechanism was exercised once on-orbit with success.

Thermoelastic effects during shadow–sunlight transitions along the orbit induce a boom shift, displacing the image of the celestial source by about 1 arcmin. The IXPE team therefore developed and implemented a correction in the flight processing pipeline. For point-like sources, automatic centroiding and position rescaling provide sufficient accuracy. For extended sources, a different strategy was required: we modeled the shift using simultaneous measurements from the two star trackers (forward and aft), temperatures measured on the boom and on the optics, and, for calibration, centroiding of point-like sources. The resulting model performs well, as illustrated by the improvement in the Cyg X-1 image before and after correction (see Figure 1).

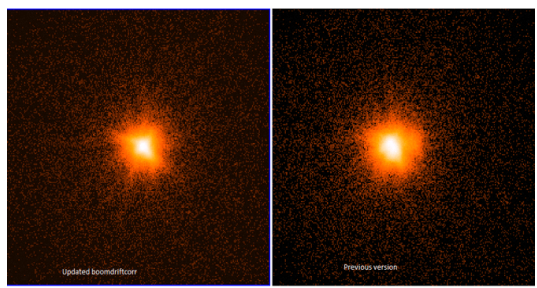

**Figure 1.** (**Left**) Cygnus X-1 image after the boom thermoelastic boom motion correction. (**Right**) Cygnus X-1 image before the boom motion correction

In fact, IXPE was eventually launched on a SpaceX Falcon 9, which offered substantially more volume and resources, but the spacecraft design had already been finalized. Given the payload volume available on Falcon 9, an improvement over the IXPE design, had this launcher been available from the start, would have been to adopt a solid, tapered carbon fiber tube in place of the deployable boom, thus eliminating the tip–tilt–rotate mechanism and mitigating thermoelastic oscillations.

## 4. Ground Calibration and the Need to Dither the Satellite's Pointing Direction

Although it is common practice in the development of a space mission to shorten the planned ground-calibration campaign and rely subsequently on on-orbit calibration, this was not possible for IXPE, a genuine discovery mission. During development, we found at low energy, where most of the photons are detected, that an unpolarized source exhibited a large spurious modulation (Figure 2 and Figure 3 a nonzero modulation in the presence of an unpolarized source, see ), far exceeding the acceptable level. At 5.9 keV, spurious modulation was found below the specified requirement, as expected. We therefore made extensive effort on the back-end electronics and on the application-specific integrated circuit (ASIC) configuration to reduce this effect to the point where the electronics contribution became negligible. Ultimately, the residual spurious modulation was reduced to a level at which the dominant contribution arose from the gas electron multiplier (GEM), about 1–2% at 2 keV. This required a final calibration campaign to characterize and filter out the remaining modulation [27]. To achieve the precision required for measuring a polarization of 1%, see Figure 4, we calibrated each detector unit (DU) at six energies, collecting 66 million counts per energy in the central part of the detector and about 42 million in the rest of the field of view, see Figure 4. Meeting the required uncertainty of 0.15% in spurious modulation consumed about 60% of the total calibration time (approximately 40 days per DU, operating 24/7).

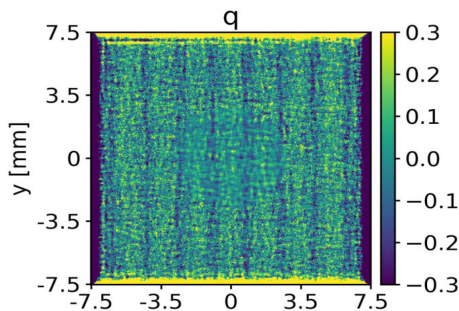

**Figure 2.** Stoke q = Q/I (2−D map). The visible strips are consistent with the laser etching pattern for producing the holes of the GEM.

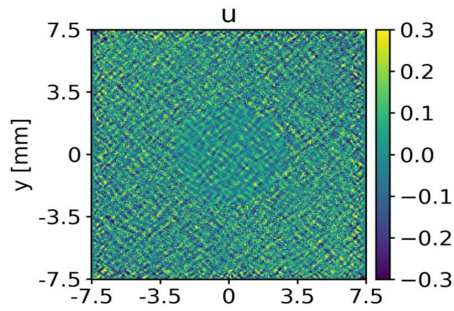

**Figure 3.** Stoke u = U/I (2−D map). The absence of definite pattern in the presence of a Q pattern indicates a coherent polarization angle across the surface.

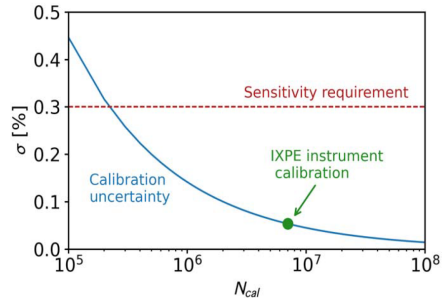

**Figure 4.** The statistical uncertainty in the measurement of the spurious modulation as a function of detected counts, its requirement, and the needed counts to meet the requirement.

Since the performance of the GEM and the ASIC, particularly the residual spurious modulation, varies across the detector on spatial scales of a few pixels, a per-pixel calibration was incompatible with the mission schedule. We therefore adopted a hybrid approach that couples calibration and operations: on the ground, we calibrated the entire field of view, but with a higher sensitivity, its central part where the most sensitive observations of bright and point-like sources were expected to be carried out; in orbit, we implemented dithering of the satellite so that each target samples many detector pixels, averaging out small-scale response variations (as done for Chandra for the same reason).

The IXPE dithering can be modeled as a two−frequency Lissajous motion with an optional slow radial term,

$$x(t) = A \sin(\omega_a t) \sin(\omega_x t), \qquad y(t) = A \sin(\omega_a t) \sin(\omega_y t), \qquad \omega_a = \frac{2\pi}{907 \,\text{s}}.$$

We adopt three selectable dither amplitudes:

$$A \in \{0.8, \ 1.6, \ 2.6\} \ \text{arcmin}$$

(small, standard—often taken as $A \simeq 1.59$ arcmin—and large). For the fast components, we use

$$\omega_x = \frac{2\pi}{101 \,\text{s}} \quad \text{and} \quad \omega_y = \frac{2\pi}{449 \,\text{s}}.$$

The period(s) of the dithering pattern are much shorter than a typical IXPE observation ($\gtrsim 10^5$ s), so the illumination across the field is effectively uniform. The dithering radius can be chosen among the small (0.8 arcmin), standard (1.6 arcmin), and large (2.6 arcmin) to be converted in the detector area considering a focal length of 4000 mm.

During calibration, we accumulate $10^6$ counts mm$^{-1}$ in the dithering region and $10^5$ counts mm$^{-1}$ outside for each energy that reaches the needed counts to meet the precision requirement in knowledge of spurious modulation. The flux rate was limited to about 200 counts s$^{-1}$ due to the 1.1 ms dead time of the IXPE ASIC, which naturally increased the total calibration time.

A practical drawback of dithering is the time-dependent vignetting of the mirror–detector system during an observation. Consequently, the effective area must be treated as a function of time,

$$A_{\text{eff}}(t) = A_0 \, V(t),$$

where $V(t)$ encodes the vignetting along the dither path.

For bright sources, residual power can appear in the data at dithering frequencies. These components should be modeled and removed, e.g., by time-dependent effective-area corrections.

## 5. Impact of GEM Charging on Gas Gain

An effect anticipated on-ground and calibrated was the process of the GEM charging during multipication of primary electron tracks.

In GEMs, the charging-up comes from charge sticking to the insulating (Kapton) walls of the holes. Electron build-up (negative charge) on the walls/rims tends to weaken/defocus the field in the hole, so the effective gain drops. Ion build-up (positive charge) usually does the opposite by strengthening/focusing the field, and the gain rises, until an equilibrium is reached (see Figure 5). The dominating one depends on fields (drift/multiplication/transfer/), hole geometry, gas, and ion backflow, but a reduction in gain is typically a signature of electron charging of the dielectric.

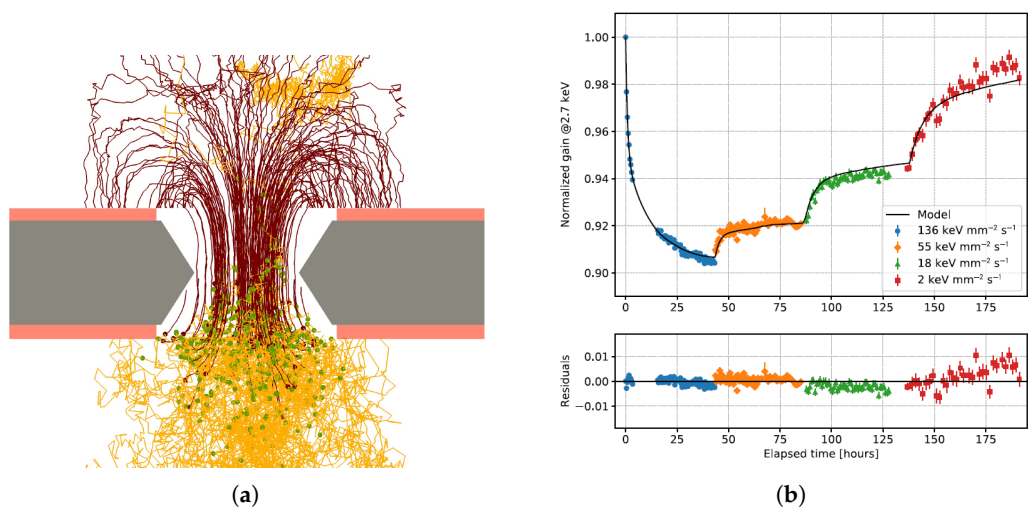

|         |         |
| :-----: | :-----: |
| (**a**) | (**b**) |

**Figure 5.** (**a**) Electrons (yellow tracks) and ions (brown tracks) produced during the multiplication process, within the GEM holes, as simulated by Garfield/Magboltz [28]. Green points are the sites of the ionization. Electrons and ions alter, with opposite sign, the electric field within the multiplication channels and above and below the exposed dielectric of the GEM. The overall impact is a decrement of the gas gain. (**b**) Example of gain drop and recover as a function of the rate [6]. If we consider the total counting rate of the Crab 80 counts/s, in a single DU [2], at an average energy of 2.7 keV, its energy density rate in the medium dithering area (1.8 mm radius) is 20 keV mm$^{-2}$ s$^{-1}$.

The use of the filter and calibration wheel (FCW), particularly its flood X-ray sources at 5.89 and 1.7 keV, mitigates this effect. Detector calibrations are routinely performed while the celestial target is occulted by the Earth. Nevertheless, the mismatch between the energy density rates of the calibration sources and those of the astrophysical targets prevents an exact compensation of the charging-induced gain variations (see Figure 6).

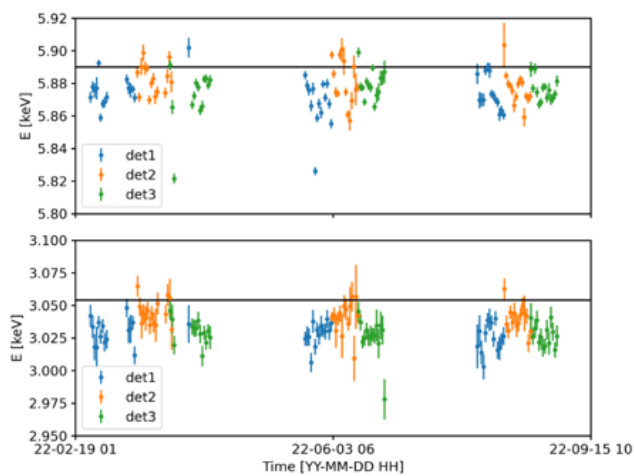

**Figure 6.** The reconstructed energy of the two on-board polarized lines (we call this source Cal A) at 5.89 keV and 3.05 keV, after calibration with on-board unpolarized and flood 5.89 (called Cal C) and 1.7 keV (called Cal D) sources. The black horizontal line represents the expected (nominal) energy of the two polarized lines.

In the flight pipeline that provides data reduction, a correction based on the modeling of charging is provided as in Equation (1).

$$\frac{dq(t)}{dt} = \overbrace{R\alpha_c \left(1 - \frac{q(t)}{q_{\max}}\right)}^{\text{charge}} - \overbrace{\alpha_d q(t)}^{\text{discharge}} \tag{1}$$

This result is directly related to the theory of capacitor charging and discharging, governed by two time constants $\alpha_c$ and $\alpha_d$. In ref. [6], a complete treatment and a closed-form expression derived from Equation (1), reparametrized in terms of nondegenerate physical constants that can be determined from ground calibration, are provided.

Unfortunately, charging could be calibrated for only two of the three flight detectors. Moreover, the simple model adopted for the gain variation, together with the on-board calibration procedure, which was not performed simultaneously with the observations, is not sufficiently accurate. As a result, the spectrum of a particularly bright source (at ∼crab flux) appears to be steeper % than expected in the absence of charging.

## 6. Absorption of Dimethyl Ether: Consequences and Mitigation Strategies

Figure 7 shows the quantum efficiency (QE) measurements of the GPDs in flight at 2.69 and 6.4 keV, obtained in three campaigns: acceptance tests, DU calibration, and Delta-DU calibration after filter and calibration wheel rework. Measurements were carried out with high accuracy using a well-calibrated commercial silicon drift detector with a thin beryllium window (by Amptek, Bedford, USA) , as well as by carefully determining the transmittance of the residual air column from up-to-date environmental parameters and measured source–detector distances.

We found that the expected quantum efficiency (QE) was not achieved for a GPD filled with 800 mbar dimethyl ether ($(CH_3)_2O$, DME); instead, the QE decreased over time. A time-dependent analysis of the energy resolution showed no degradation in performances, whereas the gas gain exhibited a clear upward trend.

$$G_{\text{eff}}^{\text{corr}} = \frac{G_{\text{eff}}}{\exp\left(C \times \left(\frac{1}{P/T} - \frac{1}{2.533}\right)\right)} \tag{2}$$

The gain increase, which is expected if the pressure drops at constant temperature (see [29] and Equation (2)), is accompanied by longer tracks with a corresponding increase in the modulation factor of the polarimeter (see Figure 8). Independent measurements at INFN–Pisa of the beryllium window curvature induced by internal pressure showed that GPDs exhibiting larger relative gain increases also displayed larger curvature. The observed curvature is well reproduced by simulations in which the internal pressure is inferred using the track length and modulation factor as proxies.

Modeling the time dependence on the pressure [6]:

$$p(t; \tau, \Delta_p) = p_0 - \Delta_p \left( 1 - \exp\left\{ -\frac{(t - t_0)}{\tau} \right\} \right) \tag{3}$$

and the dependence of track length from pressure as

$$L(p) = L_0 \left( \frac{p}{p_0} \right)^{-\alpha_L} \tag{4}$$

From a simultaneous fit of the on-board data from the calibration sources, it was possible to derive the expected efficiency at a given time.

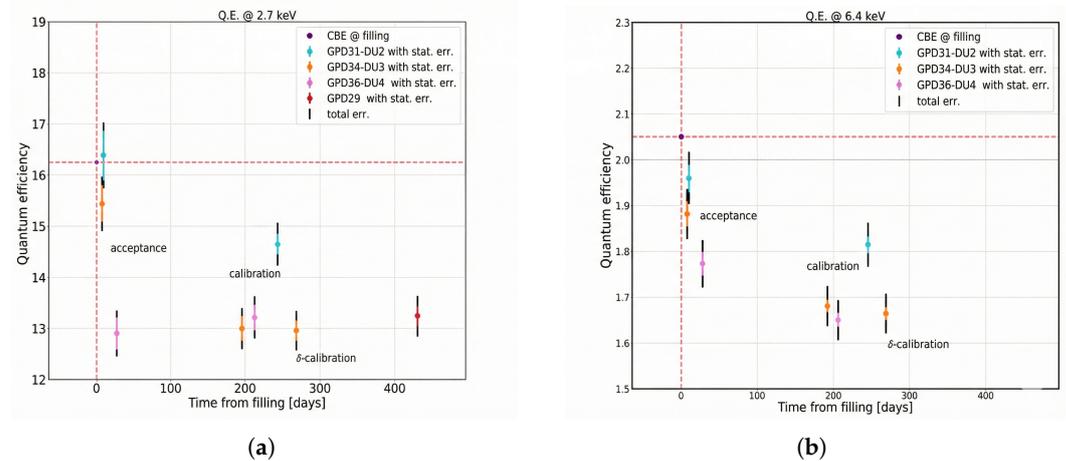

**Figure 7.** (**a**) The quantum efficiency (in %) of the flight detectors measured at three epochs at 2.69 keV. The red horizontal line corresponds to the expected value. (**b**) The quantum efficiency (in %) of the flight detectors measured at three epochs at 6.4 keV. The red horizontal line corresponds to the expected value after nominal 800 mbar filling. Data shows a decreasing trend in the quantum efficiency

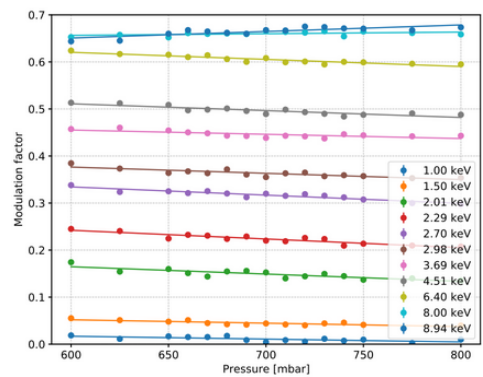

**Figure 8.** Increase of the modulation factor of the GPD as a function of the decreasing pressure obtained by Monte Carlo Simulation.

All of the evidence discussed above points to the secular absorption or adsorption of DME by one or more components of the GPD, without any associated leak (otherwise, the energy resolution would have worsened during ground tests).

The GPD is composed of several elements made from different materials and bonded together with glue. To identify which component could be responsible for DME uptake and the resulting decrease in internal pressure, we designed a dedicated laboratory experiment. We prepared six stainless-steel vessels, each filled with pure DME or with DME plus a single GPD element. Immediately after filling the vessels to exactly 800 mbar, we directly measured the internal pressure and the gas temperature. These quantities were then monitored for several months.

All vessels showed a constant DME content over time except one. The key result is shown in Figure 9, where the DME pressure inside the vessel containing only the glue is plotted. The data exhibit a rapid decay (exponential time constant of ~18 h) followed by a much slower decay (~1490 h). This clearly demonstrates that the glue is responsible for absorbing DME. The control vessel, filled only with DME, showed no pressure drop.

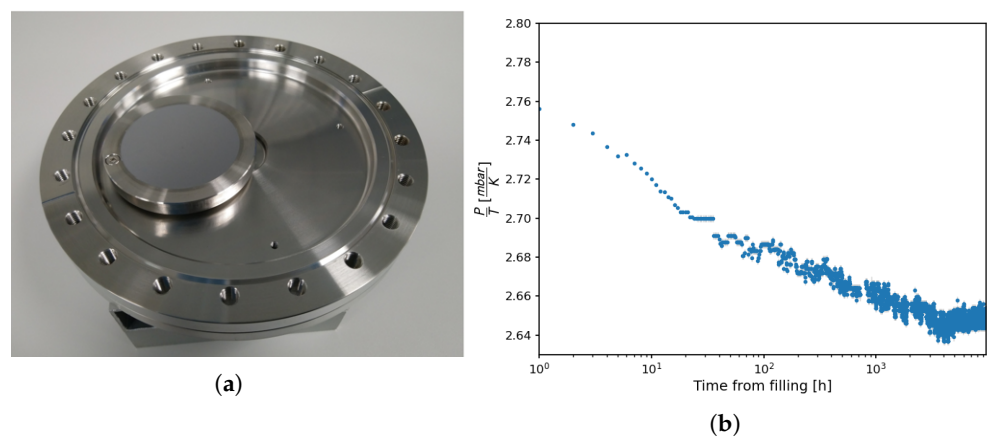

**Figure 9.** (**a**) Stainless-steel vessel, manufactured at INAF-IAPS opened to show the small container of the glue Supreme 10 HT by Master Bond, Hackensack (NJ), USA Uniti) routinely used to seal the different elements of the GPD. After closing, the vessel is filled with DME at 800 mbar. (**b**) Measured pressure trend in the closed vessel filled with DME and containing the glue shown in panel (**a**). The data exhibit a rapid pressure decay with an exponential time constant of 18 h, followed by a slower decay with a time constant of 1490 h. No such pressure decrease was observed when repeating the test with the other GPD elements or with the control vessel.

Following the test conducted under INAF–IAPS supervision at Oxford Instruments Technologies Oy (Espoo, Finland), the company responsible for filling the GPDs with flight gas mixtures, an additional test at INFN showed that accurately modeling the time evolution of the continuous pressure decrease requires a constant in-orbit loss term of about 15 mbar per year. The behavior is consistent with not only adsorption leading to saturation, but also absorption that produces a steadily decreasing pressure [30]. The results at INFN–Pisa showed better modeling of the trend using a stretched exponential as in Equation (5). Such modeling is typically used in absorption processes in solids.

$$p\left(t; p_0, \Delta_p, \alpha, \tau\right) = p_0 - \Delta_p \left(1 - \exp\left[-\left(\frac{t}{\tau}\right)^{\alpha}\right]\right) \tag{5}$$

To clarify the behavior in-orbit, we show Figure 10. The photoelectron track lengths produced by 5.89 keV photons from the on-board calibration source increase steadily with time. A corresponding steady increase in the modulation factor is observed. The length of the track is a good proxy for pressure. The modulation factor increases at all energies, as shown in Figure 8; however, the in-orbit consistency check is performed using the on-board calibration source. This behavior requires lowering the gas gain twice per year and updating the response matrices to reflect the revised quantum efficiency and modulation factor, ensuring an accurate interpretation of the results.

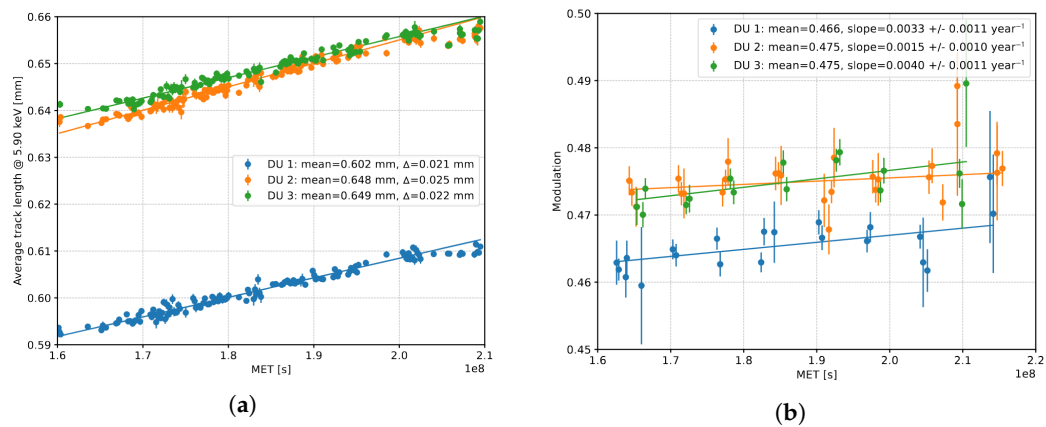

(a)                                                              (b)

**Figure 10.** (**a**) Track length from the on-board calibration source as a function of Mission Elapsed Time (MET, in seconds) since launch on 9 December 2021 at 06:00 UTC. (**b**) Corresponding modulation factor as a function of time, measured on-orbit with the on-board calibration sources.

## 7. Comparative Inefficiency of Background Rejection

The background in proportional counters has been studied and characterized for many decades, across different configurations and orbits (for a review, see [31]). Minimum ionizing particles (MIPs) deposit in gas at about 2 MeV cm$^2$ g$^{-1}$, thus producing signals within the operating energy range of these detectors (in contrast to solid-state detectors with depletion layers of ~few hundred µm in silicon). Background suppression typically combines three standard techniques: low- and high-pulse height thresholds, front/back/side anticoincidence, and pulse-shape discrimination.

The effectiveness of the last two techniques strongly depends on the detector geometry. For example, the Charpak design [32] is well suited to pulse-shaped discrimination and to implement anticoincidence systems [33,34]. The complementary PCA-style design [35] is very effective for anticoincidence systems but less suitable for pulse-shape discrimination.

When the new photoelectric X-ray polarimeters were conceived [7,8,24,36], it was soon recognized that given the large number of counts required by polarimetry (and hence the typically bright celestial sources), the instrumental background would not be the limiting factor. The early proposals therefore benchmarked the expected residual background (i.e., after all rejection techniques) to that measured with the Wisconsin experiment on board OSO-8 [37], particularly its Neon–CO$_2$ section. In the GPD case, with true two-dimensional imaging of the photoelectron track, pulse-shape discrimination was not possible because the information about the time of arrival of each pixel was lost.

However, detailed pre-flight GEANT4 simulations [38] were less encouraging: the estimated residual background for the GPD, in units of counts s$^{-1}$ cm$^{-2}$ keV$^{-1}$, was predicted to be ~10–20 times higher than that for the Wisconsin experiment (see Table 2). The origin of the unrejected component becomes clear on inspection: MIPs traversing nearly parallel to the sensitive plane distribute their ionization over many pixels; with the modest gas gain (<200), the self-trigger is not fired unless a keV-scale δ-ray is produced, which is indistinguishable from a photoelectron of the same energy.

The discrepancy between the observed and simulated background is rather small, ultimately close to a factor of ~two, once the different background-rejection efficiencies (and corresponding X-ray acceptances) applied in orbit and assumed in the ground simulations are taken into account.

**Table 2.** Residual background levels in proportional counters and GPDs (all in counts $s^{-1}$ $cm^{-2}$ $keV^{-1}$).

| Instrument/Gas | Energy Band (keV) | Residual Background |
|---|---|---|
| Wisconsin OSO-8 [37]/Ne–$CO_2$ | 1.6–3.0 | $1.5 \times 10^{-4}$ |
| Wisconsin OSO-8 [37]/Ne–$CO_2$ | 3.0–6.0 | $1.0 \times 10^{-4}$ |
| IXPE GPD/DME ([38], est $^\dagger$.) | 2.0–8.0 | $1.9 \times 10^{-3}$ |
| IXPE GPD/DME ([39], meas. $^\ddagger$) | 2.0–8.0 | $6.7 \times 10^{-3}$ |

$^\dagger$ Includes a 75% background-rejection efficiency corresponding to an X-ray acceptance of *sim*80%. $^\ddagger$
Converted from a measured surface brightness of 0.003 *counts* $s^{-1}$ $arcmin^{-2}$ (2–8 keV) to counts $s^{-1}$ $cm^{-2}$ $keV^{-1}$
using IXPE's plate scale $\simeq 51.6''$ $mm^{-1}$ (4 m focal length), i.e., 1 $arcmin^2 \simeq 1.352 \times 10^{-2}$ $cm^2$ on the focal plane,
and dividing by the 6 keV bandpass; a 40% background-rejection factor is applied while keeping the X-ray
acceptance $\gtrsim 99\%$ so as not to alter the response matrices.

In particular, it is now clear why the GPD does not achieve the same residual background as multi-wire proportional counters: (i) the absence of an anticoincidence system; (ii) the relatively low gas gain, which limits the effectiveness of background–X-ray discrimination based on track pixel multiplicity and morphology.

## 8. Dead-Time Fraction in IXPE ASICs and Techniques for Observing Bright Sources

The dead time per-event of the IXPE detector, as measured on the ground, is $\tau \simeq$ 1.1–1.2 ms. For a non-paralyzable system, the dead-time fraction for each detector unit (DU) is given by Equation (6). Observing the Crab Nebula yields a total rate of about $R_{\text{true}}^{(\text{tot})} \approx 240$ $s^{-1}$ across the three DUs [2], i.e., $R_{\text{true}}^{(\text{DU})} \approx 240/3 \approx 80$ $s^{-1}$ per DU (assuming a uniform distribution). Substituting into Equation (6) gives $f_{\text{dead}} \approx 0.081$–0.088 (i.e., $\sim$8–9% per DU).

$$f_{\text{dead}} = \frac{\tau R_{\text{true}}}{1 + \tau R_{\text{true}}} . \tag{6}$$

Most of the dead-time budget arises from (i) the acquisition of the X-ray event and the subsequent on-the-fly pedestal acquisition in the same region of interest and (ii) an overhead related to the peculiar shape of the internal analog signal. An improved version of the ASIC is expected to have a dead time $\sim 5\times$ smaller [40], which would reduce the dead-time fraction of crab per-DU to $\sim$1.7–1.9%.

Given the presence of persistent very bright X-ray sources (e.g., Sco X-1 [41]) and unanticipated transients [42] with rates $\sim$5–10$\times$ the Crab, the resulting dead-time fraction without mitigation would increase to $\sim$30–50% per DU for the current ASIC parameters. To keep the observed rate at a manageable level, FCWs [9] include an aluminized Kapton filter that attenuates the flux. This filter, used in the observations cited above, was accurately calibrated on the ground and its transmission is included in the appropriate response matrices for these observations [2].

For completeness, the forward (non-paralyzable) relation between true and observed rates is as follows:

$$R_{\text{obs}} = \frac{R_{\text{true}}}{1 + \tau R_{\text{true}}} , \tag{7}$$

which is the expression used to predict and control the recorded rate when applying flux attenuation by dead-time.

In fact, the indication for the operation of IXPE is to use the Kapton filter on the whole instrument when the observed source is brighter than twice the rate expected from the Crab Nebula.

## 9. Discussion

It is widely recognized that the Imaging X-ray Polarimetry Explorer (IXPE) has opened a new window in astrophysics: For the first time, it has enabled the detemination of geometrical parameters in compact objects without the necessity of an image at resolution not yet achievable by current or near future X-ray optics, mapped magnetic-field geometry and turbulence levels at particle-acceleration sites in extended sources, and constrained the emission mechanisms and their respective contributions to the observed polarization. Nevertheless, the instrument technology flown on IXPE entails several limitations:

1. Narrow energy band. Although broader than Bragg polarimeters, IXPE's passband remains limited compared with modern X-ray spectroscopy missions.
2. No full 3D track reconstruction. Only the 2D projection of the photoelectron track in the detector plane is available. A true 3D reconstruction would help disentangle large-angle Rutherford scattering from apparent vertical evolution of the track.
3. Large dead time. The Gas Pixel Detector (GPD) dead time necessitates the use of flux attenuators for bright sources, reducing sensitivity and complicating calibration.
4. Secular pressure drop. Due to the slow decrease in gas pressure, the detection efficiency decreases by about 2.3% $yr^{-1}$ (relative), while the modulation factor increases by 0.3% $yr^{-1}$ (absolute) (measured at 5.9 keV by the calibration source).
5. Background discrimination. Limited capability to reject minimum ionizing particles (MIPs) can hamper background suppression in high particle-background environments.
6. GEM charging. Despite the calibration source on board, ground calibrations, and associated modeling, residual (uncompensated) gain drops in the GEM can distort the spectral shape for sources as bright as $\sim$1–2 $\times$ $10^{-8}$ erg cm$^{-2}$ s$^{-1}$ unless corrected in the analysis.

These limitations were addressed through dedicated procedures and on-board subsystems, which successfully mitigated their impact on the analysis of celestial-source data. The narrow energy band, in particular, was alleviated by multi-satellite campaigns involving *NICER*, *Chandra*, *XMM*, *Swift*, *NuSTAR* , together with a spectropolarimetric approach. Moreover, extensive use of the on-board calibration system—e.g., regular exposures with the built-in source—was instrumental in tracking and correcting gain drifts and modulation factor variations.

The main limitations at spacecraft/mission level are as follows:

1. Earth occultation duty cycle. Because many targets lie in the Galactic plane, Earth occultation in a low-Earth orbit can remove roughly 50% of the available observing time.
2. Boom metrology. The use of an extension boom is not accompanied by an active metrology system to correct for thermoelastic expansion and flexure, which can degrade absolute imaging and polarization systematics.
3. Limited on-board storage memory and S-band for data transmission. Limited storage (6 GB) prevents the observer from observing bright sources for a long time. The typical S-band download rate (2 Mbit/s) requires 7 days to transmit to the ground station (Malindi) 75 ks of net observation of Crab data (flux $\sim$ 2 $\times$ $10^{-8}$ erg cm$^{-2}$ s$^{-1}$).
4. Slow response to Target of Opportunity. IXPE Mission Operation Control and IXPE spacecraft was not designed for Swift-like follow-up. The SMEX character of IXPE limits a repointing to not less than 3 working days from the reception of a ToO request https://ixpe.msfc.nasa.gov/for_scientists/too.html, accessed on 16 December 2015.

## 10. Conclusions

IXPE has been a major success. Teams from NASA, BAE, ASI, INAF, and INFN, with contributions from OHB Italia, were instrumental in resolving issues throughout the development and operational life of the satellite. Although selected as a SMEX, IXPE has served as a pathfinder mission; indeed, thanks in part to the strong Italian contribution, NASA suggested that the mission could be regarded as effectively MIDEX-class. In fact, some of us are already thinking about future developments based on a possible wide-band observatory and equipped with fast repointing capabilities (see [43]) .

Martin C. Weisskopf, Paolo Soffitta, and the IXPE team were awarded the 2024 Bruno Rossi Prize of the American Astronomical Society; Enrico Costa, Ronaldo Bellazzini and Martin C. Weisskopf received the 2025 Feltrinelli International Prize of the Accademia Nazionale dei Lincei (Italy); Enrico Costa in 2025 received the Ernst Mach Medal of the Czech Academy of Sciences (CAS) for physical science for his pioneering contribution to the field of X-ray Polarimetry in astronomy.

**Author Contributions:** Conceptualization: P.S., E.C., F.M., S.F., Software: F.M., S.F., R.F., A.D.M., F.L.M., A.T., investigation: F.M., S.F., R.F., A.D.M., F.L.M.; writing original draft preparation: PS, writing review and editing P.S., E.C., E.D.M., A.D.M., S.F., R.F., F.L.M., F.M, A.R., A.T.

**Funding:** The US contribution is supported by the National Aeronautics and Space Administration (NASA) and led and managed by its Marshall Space Flight Center (MSFC), with industry partner Ball Aerospace (now, BAE Systems). The Italian contribution is supported by the Italian Space Agency (Agenzia Spaziale Italiana, ASI) through contract ASI-OHBI-2022-13-I.0, agreements ASI-INAF-2022-19-HH.0 and ASI-INFN-2017.13-H0, and its Space Science Data Center (SSDC) with agreements ASI-INAF-2022-14-HH.0 and ASI-INFN 2021-43-HH.0, and by the Istituto Nazionale di Astrofisica (INAF) and the Istituto Nazionale di Fisica Nucleare (INFN) in Italy.

**Data Availability Statement:** IXPE Astrophysical, flight calibration and satellite ancillary data are available at HEASARC https://heasarc.gsfc.nasa.gov/docs/ixpe/archive/, ground calibration, laboratory and ancillary instrument's ground-data are available on reasonable requests.

**Acknowledgments:** The Imaging X-ray Polarimetry Explorer (IXPE) is a joint NASA and ASI mission. This research partially used data products provided by the IXPE Team (MSFC, SSDC, INAF, and INFN) and distributed with additional software tools by the High-Energy Astrophysics Science Archive Research Center (HEASARC), at NASA Goddard Space Flight Center (GSFC). The authors acknowledge contributions from INFN, OHB-Italy, and NASA-MSFC.

**Conflicts of Interest:** The authors declare no conflicts of interest.

## Abbreviations

The following abbreviations are used in this manuscript:

| AGN | Acrtive Galactic Nuclei |
| BH | Black Holes |
| DME | Dimethyl Ether |
| FCW | Filter and Calibration Wheel |
| GEM | Gas Electron Multiplier |
| GPD | Gas Pixel Detector |
| GOP | General Observer Program |
| IXPE | Imaging X-Ray Polarimetry Explorer |
| MIP | Minimum Ionizing Particle |
| NS | Neutron Star |
| OSO | Orbiting Solar Observatory |
| SMEX | Small Explorer Program |
| SNR | Supernova Remnants |
| SOC | Science Operation Center |
| QE | Quantum Efficiency |
| WD | White Dwarfs |

# References

1. Weisskopf, M.C.; Soffitta, P.; Baldini, L.; Ramsey, B.D.; O'Dell, S.L.; Romani, R.W.; Matt, G.; Deininger, W.D.; Baumgartner, W.H.; Bellazzini, R.; et al. The Imaging X-ray Polarimetry Explorer (IXPE): Pre-Launch. *J. Astron. Telesc. Instrum. Syst.* **2022**, *8*, 026002. https://doi.org/10.1117/1.JATIS.8.2.026002.

2. Soffitta, P.; Baldini, L.; Bellazzini, R.; Costa, E.; Latronico, L.; Muleri, F.; Del Monte, E.; Fabiani, S.; Minuti, M.; Pinchera, M.; et al. The Instrument of the Imaging X-Ray Polarimetry Explorer. *Astron. J.* **2021**, *162*, 208. https://doi.org/10.3847/1538-3881/ac19b0.

3. Novick, R. Stellar and Solar X-Ray Polarimetry. *Space Sci. Rev.* **1975**, *18*, 389. https://doi.org/10.1007/BF00212912.

4. Weisskopf, M.C.; Silver, E.H.; Kestenbaum, H.L.; Long, K.S.; Novick, R. A precision measurement of the X-ray polarization of the Crab Nebula without pulsar contamination. *Astroph. J.* **1978**, *220*, L117. https://doi.org/10.1086/182648.

5. Ramsey, B.D.; Bongiorno, S.D.; Kolodziejczak, J.J.; Kilaru, K.; Alexander, C.; Baumgartner, W.H.; Breeding, S.; Elsner, R.F.; Le Roy, S.; McCracken, J.; et al. Optics for the imaging x-ray polarimetry explorer. *J. Astron. Telesc. Instrum. Syst.* **2022**, *8*, 024003. https://doi.org/10.1117/1.JATIS.8.2.024003.

6. Baldini, L.; Barbanera, M.; Bellazzini, R.; Bonino, R.; Borotto, F.; Brez, A.; Caporale, C.; Cardelli, C.; Castellano, S.; Ceccanti, M.; et al. Design, construction, and test of the Gas Pixel Detectors for the IXPE mission. *Astropart. Phys.* **2021**, *133*, 102628. https://doi.org/10.1016/j.astropartphys.2021.102628.

7. Costa, E.; Soffitta, P.; Bellazzini, R.; Brez, A.; Lumb, N.; Spandre, G. An efficient photoelectric X-ray polarimeter for the study of black holes and neutron stars. *Nature* **2001**, *411*, 662–665 . https://doi.org/

8. Bellazzini, R.; Spandre, G.; Minuti, M.; Baldini, L.; Brez, A.; Cavalca, F.; Latronico, L.; Omodei, N.; Massai, M.M.; Sgro', C.; et al. Direct reading of charge multipliers with a self-triggering CMOS analog chip with 105 k pixels at 50 μm pitch. *Nucl. Instrum. Methods Phys. Res. Sect. A* **2006**, *566*, 552. https://doi.org/10.1016/j.nima.2006.07.036.

9. Ferrazzoli, R.; Muleri, F.; Lefevre, C.; Morbidini, A.; Amici, F.; Brienza, D.; Costa, E.; Del Monte, E.; Di Marco, A.; Di Persio, G.; et al. In-flight calibration system of imaging x-ray polarimetry explorer. *J. Astron. Telesc. Instrum. Syst.* **2020**, *6*, 048002. https://doi.org/10.1117/1.JATIS.6.4.048002.

10. Riegler, G.R.; Garmire, G.P.; Moore, W.E.; Stevens, J.C. A low-energy X-ray polarimeter. *Bull. Am. Phys. Soc.* **1970**, *15*, 635.

11. Sanford, P.W.; Cruise, A.M.; Culhane, J.L. Techniques for Improving the Sensitivity of Proportional Counters Used in X-Ray Astronomy. In *Non-Solar X- and Gamma-Ray Astronomy*; Gratton, L., Ed.; 1970 D. Reidel Pub. Co., Dordrecht-Holland, The Netherlands; Volume 37, p. 35.

12. Hayashida, K.; Tanaka, S.; Tsunemi, H.; Hashimoto, Y.; Ohtani, M. Optimization of polarimetry sensitivity for X-ray CCD. *Nucl. Instrum. Methods Phys. Res. Sect. A* **1999**, *436*, 96–101. https://doi.org/10.1016/S0168-9002(99)00604-X.

13. Austin, R.A.; Minamitani, T.; Ramsey, B.D. Development of a hard x-ray imaging polarimeter. In Proceedings of the X-Ray and Ultraviolet Polarimetry, San Diego, CA, USA, 28–29 July 1994; ; Society of Photo-Optical Instrumentation Engineers (SPIE) Conference Series; Volume 2010, pp. 118–125. https://doi.org/10.1117/12.168571.

14. La Monaca, A.; Costa, E.; Soffitta, P.; di Persio, G.; Manzan, M.; Martino, B.; Patria, G.; Cappuccio, G.; Zema, N. A new photoelectron imager for X-ray astronomical polarimetry. *Nucl. Instrum. Methods Phys. Res. Sect. A* **1998**, *416*, 267–277. https://doi.org/10.1016/S0168-9002(98)00486-0.

15. Sakurai, H.; Gunji, S.; Tokanai, F.; Maeda, T.; Saitoh, N.; Ujiie, N. Photoelectron track image of capillary gas proportional counter. *Nucl. Instrum. Methods Phys. Res. Sect. A* **2003**, *505*, 219–222. https://doi.org/10.1016/S0168-9002(03)01056-8.

16. Phan, N.S.; Lee, E.R.; Loomba, D. Imaging $^{55}$Fe electron tracks in a GEM-based TPC using a CCD readout. *J. Instrum.* **2020**, *15*, P05012. https://doi.org/10.1088/1748-0221/15/05/P05012.

17. Fiorina, D.; Baracchini, E.; Dho, G.; Soffitta, P.; Torelli, S.; Marques, D.J.G.; Di Giambattista, F.; Prajapati, A.; Costa, E.; Fabiani, S.; et al. HypeX: High yield polarimetry experiment in x-rays. In Proceedings of the X-Ray, Optical, and Infrared Detectors for Astronomy XI, ; Society of Photo-Optical Instrumentation Engineers (SPIE) Conference Series; Volume 13103, p. 1310318. https://doi.org/10.1117/12.3021559.

18. Tsunemi, H.; Hayashida, K.; Tamura, K.; Nomoto, S.; Wada, M.; Hirano, A.; Miyata, E. Detection of X-ray polarization with a charge coupled device. *Nucl. Instrum. Methods Phys. Res. Sect. A* **1992**, *321*, 629–631. https://doi.org/10.1016/0168-9002(92)90075-F.

19. Kotthaus, R.; Buschhorn, G.; Rzepka, M.; Schmidt, K.H.; Weinmann, P.M. Hard x-ray polarimetry exploiting directional information of the photoeffect in a charge coupled device. In Proceedings of the X-Ray and Ultraviolet Spectroscopy and Polarimetry I ; Society of Photo-Optical Instrumentation Engineers (SPIE) Conference Series; Volume 3443, pp. 105–116. https://doi.org/10.1117/12.333604.

20. Hill, J.E.; Holland, A.D.; Castelli, C.M.; Short, A.D.; Turner, M.J.; Burt, D. Measurement of x-ray polarization with small-pixel charge-coupled devices. In Proceedings of the EUV, X-Ray, and Gamma-Ray Instrumentation for Astronomy VIII, ; Society of Photo-Optical Instrumentation Engineers (SPIE) Conference Series; Volume 3114, pp. 241–249. https://doi.org/10.1117/12.283771.

21. Iwata, T.; Hagino, K.; Odaka, H.; Tamba, T.; Ichihashi, M.; Kato, T.; Ishiwata, K.; Kuramoto, H.; Matsuhashi, H.; Arai, S.; et al. Development of the X-ray polarimeter using CMOS imager: Polarization sensitivity of a 1.5 μm pixel CMOS sensor. *Nucl. Instrum. Methods Phys. Res. Sect. A* **2024**, *1065*, 169487. https://doi.org/10.1016/j.nima.2024.169487.

22. Soffitta, P.; Costa, E.; Morelli, E.; Bellazzini, R.; Brez, A.; Raffo, R. Sensitivity to x-ray polarization of a micragap gas proportional counter. In Proceedings of the X-Ray and EUV/FUV Spectroscopy and Polarimetry, ; Society of Photo-Optical Instrumentation Engineers (SPIE) Conference Series; Volume 2517, pp. 156–163. https://doi.org/10.1117/12.224922.

23. Soffitta, P.; Costa, E.; di Persio, G.; Morelli, E.; Rubini, A.; Bellazzini, R.; Brez, A.; Raffo, R.; Spandre, G.; Joy, D. Astronomical X-ray polarimetry based on photoelectric effect with micragap detectors. *Nucl. Instrum. Methods Phys. Res. Sect. A* **2001**, *469*, 164–184. https://doi.org/10.1016/S0168-9002(01)00772-0.

24. Bellazzini, R.; Spandre, G.; Minuti, M.; Baldini, L.; Brez, A.; Latronico, L.; Omodei, N.; Razzano, M.; Massai, M.M.; Pesce-Rollins, M.; et al. A sealed Gas Pixel Detector for X-ray astronomy. *Nucl. Instrum. Methods Phys. Res. Sect. A* **2007**, *579*, 853. https://doi.org/10.1016/j.nima.2007.05.304.

25. Sauli, F. GEM: A new concept for electron amplification in gas detectors. *Nucl. Instrum. Methods Phys. Res. Sect. A* **1997**, *386*, 531–534. https://doi.org/10.1016/S0168-9002(96)01172-2.

26. Tamagawa, T.; Hayato, A.; Iwahashi, T.; Konami, S.; Fumi, A. Ideal gas electron multipliers (GEMs) for X-ray polarimeters. In *X-ray Polarimetry: A New Window in Astrophysics by Ronaldo Bellazzini, Enrico Costa, Giorgio Matt and Gianpiero Tagliaferri*; Bellazzini, R., Costa, E., Matt, G., Tagliaferri, G., Eds.; Cambridge University Press, Cambridge, UK : 2010; p. 60, ISBN: 9780521191845.

27. Rankin, J.; Muleri, F.; Tennant, A.F.; Bachetti, M.; Costa, E.; Di Marco, A.; Fabiani, S.; La Monaca, F.; Soffitta, P.; Tobia, A.; et al. An algorithm to calibrate and correct the response to unpolarized radiation of the X-ray polarimeter on board IXPE. *Astron. J.* **2022**, *163*, 39. https://ui.adsabs.harvard.edu/abs/2022AJ....163...39R.

28. Böhmer, F. V.; Ball, M.; Dørheim, S.; Höppner, C.; Ketzer, B.; Konorov, I.; Neubert, S.; Paul, S.; Rauch, J.; Vandenbroucke, M.; Space-Charge Effects in an Ungated GEM-based TPC. *arXiv* **2012**, arXiv:1209.0482.

29. Tamagawa, T.; Hayato, A.; Asami, F.; Abe, K.; Iwamoto, S.; Nakamura, S.; Harayama, A.; Iwahashi, T.; Konami, S.; Hamagaki, H.; et al. Development of thick-foil and fine-pitch GEMs with a laser etching technique. *Nucl. Instrum. Methods Phys. Res. Sect. A* **2009**, *608*, 390–396. https://doi.org/10.1016/j.nima.2009.07.014.

30. Tomaiuolo, C.; Manfreda, A.; Orsini, L.; Tugliani, S. Time-dependent instrumental effects in IXPE: Pressure variation and GEM charging inside GPDs. *Nucl. Instrum. Methods Phys. Res. Sect. A* **2024**, *1069*, 169881. https://doi.org/10.1016/j.nima.2024.169881.

31. Soffitta, P.; Campana, R.; Costa, E.; Fabiani, S.; Muleri, F.; Rubini, A.; Bellazzini, R.; Brez, A.; Minuti, M.; Pinchera, M.; et al. The background of the gas pixel detector and its impact on imaging X-ray polarimetry. In Proceedings of the Society of Photo-Optical Instrumentation Engineers (SPIE) Conference Series, San Francisco, CA, USA, 20–22 January 2012; Volume 8443. https://doi.org/10.1117/12.925385.

32. Charpak, G.; Bouclier, R.; Bressani, T.; Favier, J.; Zupančič, Č. The use of multiwire proportional counters to select and localize charged particles. *Nucl. Instrum. Methods* **1968**, *62*, 262–268. https://doi.org/10.1016/0029-554X(68)90371-6.

33. Feroci, M.; Costa, E.; Dwyer, J.; Ford, E.; Kaaret, P.; Rapisarda, M.; Soffitta, P. Background in xenon filled X-ray detectors. *Nucl. Instrum. Methods Phys. Res. Sect. A* **1996**, *371*, 538–543. https://doi.org/10.1016/0168-9002(95)01024-6.

34. Soffitta, P.; Costa, E.; Kaaret, P.; Dwyer, J.; Ford, E.; Tomsick, J.; Novick, R.; Nenonen, S. Proportional counters for the stellar X-ray polarimeter with a wedge and strip cathode pattern readout system. *Nucl. Instrum. Methods Phys. Res. Sect. A* **1998**, *414*, 218–232. https://doi.org/10.1016/S0168-9002(98)00572-5.

35. Jahoda, K.; Markwardt, C.B.; Radeva, Y.; Rots, A.H.; Stark, M.J.; Swank, J.H.; Strohmayer, T.E.; Zhang, W. Calibration of the Rossi X-Ray Timing Explorer Proportional Counter Array. *Astrophys. J. Suppl. Ser.* **2006**, *163*, 401–423. https://doi.org/10.1086/500659.

36. Black, J.K.; Baker, R.G.; Deines-Jones, P.; Hill, J.E.; Jahoda, K. X-ray polarimetry with a micropattern TPC. *Nucl. Instrum. Methods Phys. Res. Sect. A* **2007**, *581*, 755–760. https://doi.org/10.1016/j.nima.2007.08.144.

37. Bunner, A.N. Soft X-ray results from the Wisconsin experiment on OSO-8. *Astroph. J.* **1978**, *220*, 261–271. https://doi.org/10.1086/155902.

38. Xie, F.; Ferrazzoli, R.; Soffitta, P.; Fabiani, S.; Costa, E.; Muleri, F.; Marco, D. A study of background for IXPE. *Astropart. Phys.* **2021**, *128*, 102566. https://doi.org/10.1016/j.astropartphys.2021.102566.

39. Di Marco, A.; Soffitta, P.; Costa, E.; Ferrazzoli, R.; La Monaca, F.; Rankin, J.; Ratheesh, A.; Xie, F.; Baldini, L.; Del Monte, E.; et al. Handling the Background in IXPE Polarimetric Data. *Astron. J.* **2023**, *165*, 143. https://doi.org/10.3847/1538-3881/acba0f.

40. Minuti, M.; Baldini, L.; Bellazzini, R.; Brez, A.; Ceccanti, M.; Krummenacher, F.; Latronico, L.; Lucchesi, L.; Manfreda, A.; Orsini, L.; et al. XPOL-III: A new-generation VLSI CMOS ASIC for high-throughput X-ray polarimetry. *Nucl. Instrum. Methods Phys. Res. Sect. A* **2023**, *1046*, 167674. https://doi.org/10.1016/j.nima.2022.167674.

41. La Monaca, F.; Di Marco, A.; Poutanen, J.; Bachetti, M.; Motta, S.E.; Papitto, A.; Pilia, M.; Xie, F.; Bianchi, S.; Bobrikova, A.; et al. Highly Significant Detection of X-Ray Polarization from the Brightest Accreting Neutron Star Sco X-1. *Astrophys. J. Lett.* **2024**, *960*, L11. https://doi.org/10.3847/2041-8213/ad132d.

42. Veledina, A.; Muleri, F.; Dovčiak, M.; Poutanen, J.; Ratheesh, A.; Capitanio, F.; Matt, G.; Soffitta, P.; Tennant, A.F.; Negro, M.; et al. Discovery of X-Ray Polarization from the Black Hole Transient Swift J1727.8-1613. *Astrophys. J. Lett.* **2023**, *958*, L16. https://doi.org/10.3847/2041-8213/ad0781.

43. Muleri, F.; Cesare, S.; Costa, E.; Cugno, W.; Desch, K.; Di Marco, A.; Fabiani, S.; Ferrazzoli, R.; Gruber, M.; Heuchel, D.; et al. Instruments for focal plane X-ray polarimetry in the next decade. Submitted to Particles as proceeding of Advances in Space AstroParticle Physics, Sant Feliu de Guíxols, Girona (ES) (2025).